\documentclass[aps,showpacs,preprintnumbers,amsmath,amssymb,prl,twocolumn,superscriptaddress,longbibliography
]{revtex4-1}
\usepackage{dcolumn}
\usepackage{color}
\usepackage{subfigure}  
\def\comment#1{}

\usepackage{bm,latexsym,mathrsfs,enumerate,color}
\usepackage[mathcal]{euscript}
\usepackage[breaklinks=true,unicode=true,urlcolor = blue,colorlinks = true,citecolor = blue,linkcolor = blue]{hyperref}
\usepackage{graphicx}
\usepackage{todonotes}
\usepackage{wrapfig}

\renewcommand{\vec}[1]{\bm{#1}}

\def\slashchar#1{\setbox0=\hbox{$#1$}           
   \dimen0=\wd0                                 
   \setbox1=\hbox{/} \dimen1=\wd1               
   \ifdim\dimen0>\dimen1                        
      \rlap{\hbox to \dimen0{\hfil/\hfil}}      
      #1                                        
   \else                                        
      \rlap{\hbox to \dimen1{\hfil$#1$\hfil}}   
      /                                         
   \fi}                                         %

\DeclareMathAlphabet\mathbfcal{OMS}{cmsy}{b}{n}

\def\sigmab{{\mbox{\boldmath $\sigma$}}}
\def\nablab{{\mbox{\boldmath $\nabla$}}}

\begin{document}


\title{Fluctuation-induced N\'eel and Bloch skyrmions at topological insulator surfaces}

\author{Flavio S. Nogueira}
\affiliation{Institute for Theoretical Solid State Physics, IFW Dresden, Helmholtzstr. 20, 01069 Dresden, Germany}
\affiliation{Institut f{\"u}r Theoretische Physik III, Ruhr-Universit\"at Bochum,
	Universit\"atsstra\ss e 150, DE-44801 Bochum, Germany}

\author{Ilya Eremin}
\affiliation{Institut f{\"u}r Theoretische Physik III, Ruhr-Universit\"at Bochum,
	Universit\"atsstra\ss e 150, DE-44801 Bochum, Germany}
\affiliation{National University of Science and Technology “MISiS”, Moscow, 119049, Russia}

\author{Ferhat Katmis}
\affiliation{Francis Bitter Magnet Lab and Plasma Science and Fusion Center, Massachusetts 
	Institute of Technology, Cambridge, MA 02139, USA}
\affiliation{Department of Physics, Middle East Technical University, 06800 Ankara, Turkey}

\author{Jagadeesh S. Moodera}
\affiliation{Francis Bitter Magnet Lab and Plasma Science and Fusion Center, Massachusetts 
	Institute of Technology, Cambridge, MA 02139, USA}
\affiliation{Deparment of Physics, Massachusetts Institute of Technology, Cambridge, MA-02139, USA 
}

\author{Jeroen van den Brink}
\affiliation{Institute for Theoretical Solid State Physics, IFW Dresden, Helmholtzstr. 20, 01069 Dresden, Germany}
\affiliation{Institute for Theoretical Physics, TU Dresden, 01069 Dresden, Germany}

\author{Volodymyr P. Kravchuk}
\affiliation{Institute for Theoretical Solid State Physics, IFW Dresden, Helmholtzstr. 20, 01069 Dresden, Germany}
\affiliation{Bogolyubov Institute for Theoretical Physics of National Academy of Sciences of Ukraine, 03680 Kyiv, Ukraine}

\date{Received \today}

\begin{abstract}
Ferromagnets in contact with a topological insulator have become appealing candidates for spintronics due to the presence of Dirac surface states with spin-momentum locking.
Because of this bilayer Bi$_2$Se$_3$-EuS structures, for instance, show a finite magnetization at the interface at temperatures well exceeding the Curie temperature of bulk EuS. 
Here we determine theoretically the effective magnetic interactions at a topological insulator-ferromagnet interface {\it above} the magnetic ordering temperature.
We show that by integrating out the Dirac fermion fluctuations an effective Dzyaloshinskii-Moriya interaction and magnetic charging interaction emerge.
As a result individual magnetic skyrmions and extended skyrmion lattices can form at interfaces of ferromagnets and topological insulators, the first indications of which have been very recently observed experimentally.
%
%
\end{abstract}

\maketitle

{\it Introduction---} The spin-momentum locking property of three-dimensional topological insulators (TIs) \cite{Hasan10,Qi11} make them promising candidate materials for future spin-based  electronic devices. 
One important consequence of spin-momentum locking in TIs is the topological electromagnetic response, which arises from induced Chern-Simons (CS) terms \cite{Deser82,*Niemi83,*Redlich84} on each surface \cite{Qi08}.  
This happens for instance when time-reversal (TR) symmetry is broken, which renders the surface Dirac fermions gapped. 
This can be achieved, for example, by proximity-effect with a ferromagnetic insulator (FMI) 
\cite{Katmis16,Wei13,Garate10,Yokoyama10,Tserkovnyak12,Nogueira12,Nogueira13,Tserkovnyak15,Li15,Rex16,Rex16a}. 
In this case,  a CS term is generated if there are an odd number of {\it gapped} Dirac fermions, which is achieved only in the presence of out-of-plane exchange fields \cite{Nogueira13}. 
The realization of several physical effects related to the CS term that have been 
predicted in the literature critically depend on growing technologies required for the fabrication of heterostructures involving both strong TIs and FMIs. Recently, high quality Bi$_2$Se$_3$-EuS bilayer structures have been shown to exhibit proximity-induced ferromagnetism on the surface of Bi$_2$Se$_3$ \cite{Wei13,Yang13,Lee16}. 
%
%
Other successful realizations of the stable ferromagnetic TI interfaces were demonstrated recently \cite{Tang17,Zhang18}.   
%
In addition it was shown that the interface of FMI and TI can have magnetic ordering temperature much higher than the bulk ordering temperature \cite{Katmis16}, indicating that topological surface states can strongly affect the magnetic properties of a proximity-coupled FMI.

These experimental advances motivate us to investigate the effective magnetic interactions that result from the fluctuating momentum-locked Dirac fermion surface states of a TI in contact with an FMI.
We show that even in the absence of any spontaneous magnetization, at temperatures above the Curie temperature of the FMI, intriguing topologically stable magnetic textures, i.e., skyrmions, are induced as a result of quantum fluctuations of the Dirac fermions at the interface. 
In fact, we demonstrate that integrating out Dirac fermions coupled to a FMI thin film generates a Dzyaloshinskii-Moriya interaction (DMI), 
that depending on the form of the Dirac Hamiltonian, favors either N\'eel-or Bloch-type skyrmions \cite{Nagaosa13,Fert17,Wiesendanger16,Bogdanov89}. 
However, skyrmions induced in TI-FMI structures feature in addition a "charging energy'', due to the generation of a term proportional to the square of  the so called magnetic charge, $\nablab\cdot{\bf n}$, where ${\bf n}$ denotes the direction of the magnetization  
\footnote{In contrast to the weak contribution from the nonlocal energy of the volume nagnetostatic charges, which for a thin film scales quadratically with the thickness, the considered ``magnetic charging energy'' is linear with the thickness and, therefore, it can not be neglected.}. 
An important feature of our finding is that the Dirac fermions that are integrated out {\it are not gapped}, since there is no spontaneous magnetization above $T_c$ that would lead to a gap in the Dirac spectrum. Furthermore, 
the generated DMI is only nonzero if the chemical potential does not vanish.   
We obtain the phase diagram for the skyrmion solutions and identify the region of stability for skyrmion lattices in presence of the magnetic charging energy. 
This region we determine numerically by analyzing the excitation spectrum of the skyrmion solution. 
An important discovery is that the magnetic charging energy  modifies the phase diagram significantly in the case of DMIs favoring N\'eel skyrmions, the situation relevant for Bi$_2$Se$_3$-EuS interface. 
Our theoretical findings support conceptually the recent experimental observation of a skyrmion texture at a ferromagnetic heterostructure of Cr doped Sb$_2$Te$_3$~\cite{Zhang18}.  Having a skyrmion profile on a TI surface will cause significant changes in the conductance that may be 
observed in transport measurements \cite{Andrikopoulos17}. 

{\it Interface exchange interactions---} 
The Hamiltonian governing the Dirac fermions at the interface of a FMI/TI heterostructure has the general form, 
\begin{equation}
\label{Eq:Dirac-main}
H_{\rm Dirac}({\bf n}({\bf r}))=\left[{\bf d}\left(-i\hbar\nablab\right)-J_0{\bf n}({\bf r})\right]\cdot\sigmab,
\end{equation}
where ${\bf r}=(x,y)$,  
$\sigmab=(\sigma_x,\sigma_y,\sigma_z)$ is a vector of Pauli matrices and $J_0$ is the interface exchange coupling. 
The operator ${\bf d}$ is a function of the momentum operator $-i\hbar\nablab$. 
Here we consider the two possibilities leading to a Dirac spectrum, 
\begin{equation}
\label{Eq:d}
{\bf d}_1=-i\hbar v_F\nablab, ~~~~~~~~~{\bf d}_2=-i\hbar v_F\nablab\times\hat{\bf z},
\end{equation}
with the latter arising in TIs like Bi$_2$Se$_3$, Bi$_2$Te$_3$, and 
Sb$_2$Te$_3$ \cite{Zhang.PhysRevB.82.045122}. 
Experimentally, in order for the effective Hamiltonian (\ref{Eq:Dirac-main}) to give a valid low-energy description of the physics at 
the interface, the TI must be at least 7 nm thick.
The end result will be that ${\bf d}_1$ induces a DMI of the type ${\bf n}\cdot(\nablab\times{\bf n})$, which is often referred to as a  bulk DMI, but for clarity we call it {\it Bloch DMI}.
Instead ${\bf d}_2$ leads to different 
type of DMI, $\sim{\bf n}\cdot[(\hat{\bf z}\times\nablab)\times{\bf n}]=({\bf n}\cdot\nablab)\mathrm{n}_z-\mathrm{n}_z(\nablab\cdot{\bf n})$, in the magnetic literature sometimes known as  surface DMI, but to which we refer as {\it N\'eel DMI}.

The effective energy $E_{\rm eff}$ of the system is obtained by integrating out the Dirac fermions 
$c=(c_\uparrow,c_\downarrow)$ in the partition function, 
\begin{eqnarray}
\label{Eq:Partition}
&&e^{-\beta E_{\rm eff}(\vec{\mathrm{n}})}=e^{-\beta \rho_s L\int_S dS(\vec{\nabla}\vec{\mathrm{n}})^2}
\nonumber\\
&\times&\int\mathcal{D}c^\dagger\mathcal{D}ce^{-\int_{0}^{\beta}d\tau\int d^2rc^\dagger[\partial_\tau-\mu+H_{\rm Dirac}({\bf n}({\bf r}))]c},
\end{eqnarray}
where $\rho_s$ is the magnetization stiffness of the FMI, $L$ is the film thickness and the integration is over the film area $S$. 
Due to the nonzero 
$z$-component of the magnetization, the above model yields a gapped Dirac spectrum for $T<T_c$ with spin wave excitations, which 
give rise to a Chern-Simons term \cite{Nogueira12}. However, this 
gap does not occur for $T>T_c$. In the following we assume that the gap vanishes for $T\geq T_c$ and obtain the corresponding 
corrections to the free energy after integrating out the gapless Dirac fermions. 

 {\it Effective free energy and induced DMI} --- 
 The non-interacting Green function for a spin-momentum locked system can be written in general as 
 \begin{equation}\label{eq:Green-ab}
 {\cal G}_{\alpha\beta}(\omega_n,{\bf k})=G(\omega_n,{\bf k})\delta_{\alpha\beta}+{\bf F}(\omega_n,{\bf k})\cdot\sigmab_{\alpha\beta},
 \end{equation}
 where $\omega_n=(2n+1)\pi/\beta$ is the fermionic Matsubara frequency. 
From the Hamiltonian (\ref{Eq:Dirac-main}) and the functional integral in (\ref{Eq:Partition}) we have,
 \begin{equation}\label{eq:Green}
 G(\omega_n,{\bf k})=\frac{i\omega_n+\mu}{(i\omega_n+\mu)^2-{\bf d}^2({\bf k})},
 \end{equation}
 \begin{equation}\label{eq:F}
 {\bf F}(\omega_n,{\bf k})=-\frac{{\bf d}({\bf k})}{(i\omega_n+\mu)^2-{\bf d}^2({\bf k})}, 
 \end{equation}
where  ${\bf d}({\bf k})$ is either ${\bf d}_1$ or ${\bf d}_2$ from Eq. (\ref{Eq:d}) in momentum space.  
Integrating out the fermions and expanding the free energy expression up to $J_0^2$, we obtain 
after a long but straightforward calculation, the following correction to the effective free energy density \cite{Note3}
\begin{eqnarray}
\label{Eq:FE-Dirac-main}
\delta{\cal F}_{\rm Dirac}^{\rm mag}&=&\frac{s}{2}\left\{[\nablab{\bf n}({\bf r})]^2 
+[\nablab\cdot{\bf n}({\bf r})]^2\right\}
\nonumber\\
&+&i\frac{a}{2} {\bf n}({\bf r})\cdot[{\bf d}(-i\hbar\nablab)\times{\bf n}({\bf r})],
\end{eqnarray}
where $(\nablab{\bf n})^2=\sum_{i=x,y,z}(\nablab \mathrm{n}_i)^2$ defines the usual exchange term, and 
we have defined $s=\beta J_0^2/[24\pi\cosh^2(\beta\mu/2)]$ and $a=J_0^2(8\pi\hbar v_F)^{-1}\tanh(\beta\mu/2)$. 
We can drop the constant term 
$F_{\rm Dirac}(0)$ from the free energy, since it does not depend on the field. Thus, we can safely write 
${\cal F}_{\rm Dirac}=\delta{\cal F}_{\rm Dirac}$. 
The above expression features a DMI induced by Dirac fermion fluctuations. In addition, a contribution $\sim (\nablab\cdot{\bf n})^2$ 
is also generated. We will see below that the presence of this term leads to interesting physical properties when ${\bf d}_2$ is replaced for 
${\bf d}$ in Eq. (\ref{Eq:FE-Dirac-main}), modifying in this way the behavior of N\'eel skyrmions.  
Note that differently from the case where the Dirac fermion is gapped \cite{Rex16a}, no intrinsic anisotropy is generated by the Dirac fermions. 
At the same time, we note that the  form of $\delta{\cal F}_{\rm Dirac}^{\rm mag}$ including the DMI term will persist also below $T_c$ as long as the chemical potential is outside the gap, meaning that the TI surface is metallic, despite the generated mass $m$ for the Dirac 
fermions.

{\it Effective magnetic energy in an external field} --- 
The contributions from the FMI and Dirac fermions allows one to recast the effective 
energy for a thin ferromagnetic layer in the form,

\begin{figure}
	\includegraphics[width=\columnwidth]{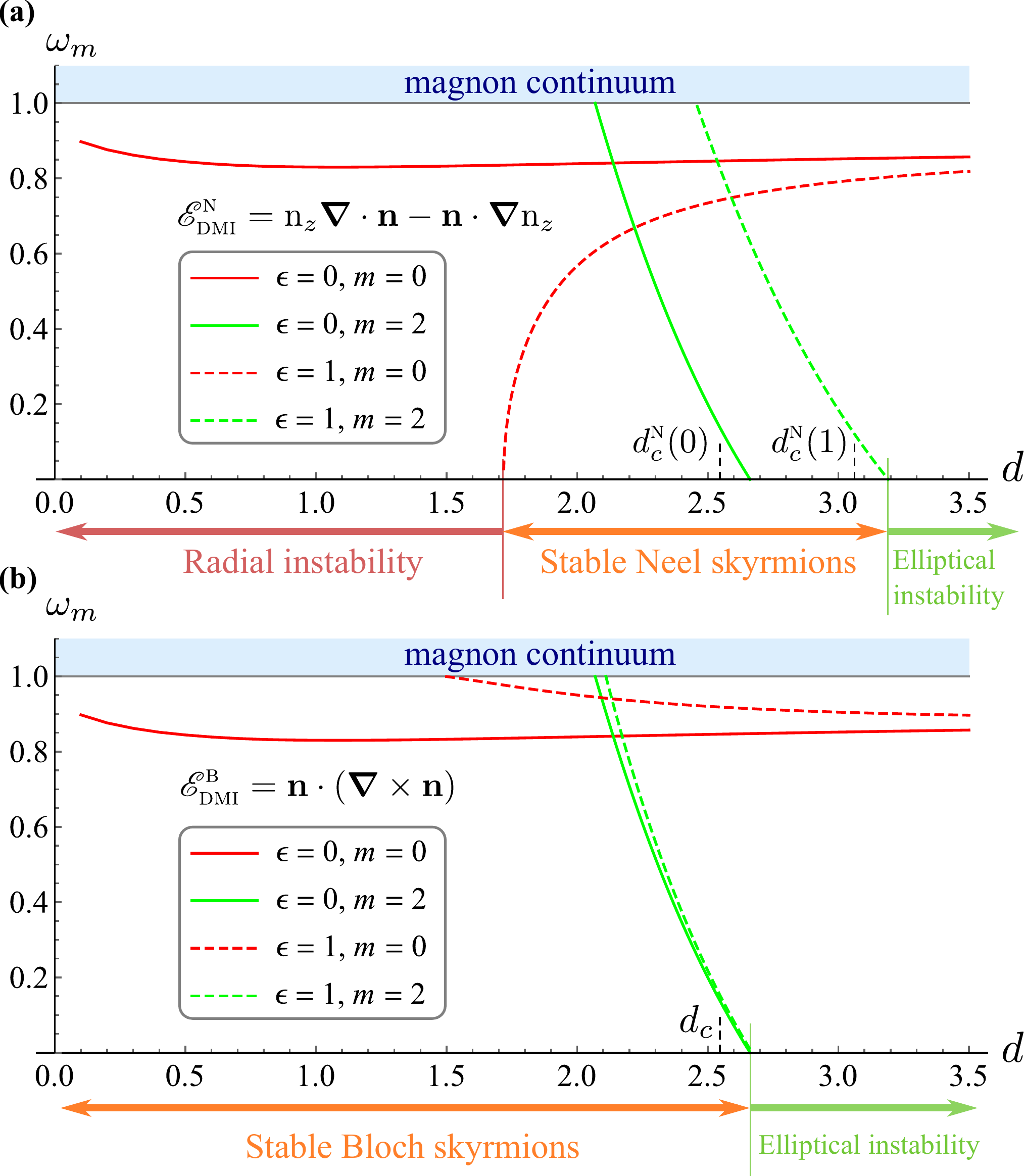}
	\caption{Eigenfrequencies of two localized modes, namely radially-symmetric ($m=0$) and elliptic ($m=2$) are found by means of numerical solution of the eigenvalue problem for different DM terms. Modes, which do not demonstrate instability, are not shown. Stability/instability regions are indicated for the case $\epsilon=1$.}\label{fig:spectrum}
\end{figure}

\begin{eqnarray}
\label{eq:E-H}
E_{\rm eff}&=&L\int\limits_S\left\{A\left[(\nablab{\bf n})^2+\epsilon(\nablab\cdot{\bf n})^2\right]\right.
\nonumber\\
&+&\left.D\mathscr{E}_\textsc{dmi}+M_sH(1-\mathrm{n}_z)\right\}\mathrm{d}S,
\end{eqnarray}
where $A=\rho_s+s/(2L)$ is the effective magnetization stiffness including the fluctuations due to the Dirac fermions. 
 We assumed 
 that the sample lies in the presence of an external magnetic field $H$ applied perpendicular to it. We have also introduced the parameter 
 $\epsilon=s/(2AL)=s/(2\rho_sL+s)$.  
 The DM coupling is given by $D=a/(2L)$. The DM interaction has the possible forms, 
 $\mathscr{E}_\textsc{dmi}^{\textsc{b}}={\bf n}\cdot(\nablab\times{\bf n})$ or $\mathscr{E}_\textsc{dmi}^{\textsc{n}}=\mathrm{n}_z\vec{\nabla}\cdot{\bf n}-{\bf n}\cdot\vec{\nabla}\mathrm{n}_z$, depending on whether $\vec{\mathrm{d}}_1$ or 
 $\vec{\mathrm{d}}_2$ arises in the Dirac Hamiltonian (\ref{Eq:Dirac-main}). The latter is more adequate for 
 Bi$_2$Se$_3$-EuS samples \cite{Li15}. The ab initio results from 
 Ref.~\onlinecite{Kim17a} indicate that $J_0$ is largely enhanced due to RKKY interactions at the Bi$_2$Se$_3$-EuS interface, ranging from 
 35 to 40 meV. Using $J_0=35$ meV one can estimate that at room temperature $s\in [0.05,0.63)$ meV, 
  and therefore $\epsilon\in[0.08,0.51)$ for 1 nm thick film and $\mu\in(0,0.1]$~eV \footnote{For EuS we would have $\rho_s=\mathcal{S}^2[J_1+J_2/2]/a_0\approx 0.29$ meV/nm, where $\mathcal{S}=7/2$ is spin of the Eu atom, $J_1/k_\textsc{b}=0.228$ K and $J_2/k_\textsc{b}=-0.118$ K are the nearest and next-nearest neighbor exchange energies, respectively, and  for the lattice spacing $a_0=5.968$ \AA~\cite{MAUGER198651}}. Note that $\epsilon$ strongly depends on the value of $\mu$, which can be reduced by doping.
 
 Although the temperature fluctuations usually destroys skyrmions in thin films, the individual skyrmions \cite{Moreau-Luchaire16,Soumyanarayanan17,Boulle16} as well as skyrmion lattices \cite{Woo16} are observed in various multilayer structures for room temperatures. Therefore, in experiments, it is reasonable to use a multilayer structure in form of the periodically repeated stack TI/FMI/NI, where NI is a normal insulator. In the following we neglect the influence of the thermal fluctuations on the magnetization structure, which holds when model \eqref{eq:E-H} is applied for a multilayer structure.
 
 Before studying the energy functional (\ref*{eq:E-H}), let us emphasize that while the DMI is absent for the case of a 
 vanishing chemical potential, the term $(\nablab\cdot{\bf n})^2$ is always there, even if $\mu=0$. 
 Thus, this term is a unique feature of thin film FMIs proximate to a three-dimensional TI. In fact, it has been recently demonstrated 
 that it is also induced for $\mu=0$ at zero temperature when the surface Dirac fermions are gapped by proximity effect to the FMI 
 \cite{Rex16a}.   

Ground states of system \eqref{eq:E-H} are well studied for the case $\epsilon=0$ \cite{Bogdanov94,Bogdanov94a,Bogdanov99,Wilson14,Schuette14}. The uniform saturation along the field is the ground state with $E_{\rm eff}=0$ for large field and weak DM interaction, and 1D structure in form of periodical sequence of $2\pi$ domain walls is the ground state with $E_{\rm eff}<0$ for small fields and strong DM interaction. The criterion for the periodical state appearance is negative energy of a single domain wall, it reads $d>d_c=8/\pi$, where $d=\sqrt{2}D/\sqrt{AM_sH}$ is dimensionless DM constant. In vicinity of the boundary $d\approx d_c$, an intermediate phase in form of 2D periodical structure (skyrmion lattice) forms the ground state \cite{Bogdanov94,Roessler06,Nagaosa13,Fert17}. An isolated skyrmion \cite{Bogdanov94a,Leonov16,Wiesendanger16,Fert17} may appear as a topologically stable excitation of the uniformly saturated state. The slyrmions and domain walls are of Bloch and N\'eel types for the DM interaction in form $\mathscr{E}_\textsc{dmi}^{\textsc{b}}$ and $\mathscr{E}_\textsc{dmi}^{\textsc{n}}$, respectively. 
 
Here we study how the ground states and individual skyrmions are changed when $\epsilon>0$. Since $\nablab\!\cdot\!{\bf n}\equiv0$ for any domain wall and skyrmion of the Bloch type (induced by $\mathscr{E}_\textsc{dmi}^{\textsc{b}}$) the influence of the term $(\nablab\!\!\cdot\!\!{\bf n})^2$ is not significant in this case. However, it drastically changes the ground state digram and stability of the static solutions for the case of $\mathscr{E}_\textsc{dmi}^{\textsc{n}}$. In this case, $d_c=d_c^{\textsc{n}}(\epsilon)=(8/\pi)\int_{0}^{1}\!\!\sqrt{1+\epsilon(2\xi^2-1)^2}\,\mathrm{d}\xi$ and period of the 1D structure is increased with $\epsilon$ \footnote{See the supplemental materials.}.  Energy per period is $E_{\textsc{1d}}^{\textsc{n}}(d,\epsilon)\approx AL\mathcal{E}(d,\epsilon)$, where $\mathcal{E}(d,\epsilon)$ is determined by the implicit relation $d/d_c^{\textsc{n}}(\epsilon)=\mathrm{E}(4/\mathcal{E})\sqrt{-\mathcal{E}/4}$, with  $\mathrm{E}(k)$ being the complete elliptic integral of the second kind \cite{Abramowitz72} (note that $\mathcal{E}<0$). For the case $\mathscr{E}_\textsc{dmi}^{\textsc{b}}$ the 1D periodical structure is not affected by $\epsilon$ and one has $d_c^{\textsc{b}}=d_c^{\textsc{n}}(0)$ and $E_{\textsc{1d}}^{\textsc{b}}(d)=E_{\textsc{1d}}^{\textsc{n}}(d,0)$ \cite{Note3}.

{\it Skyrmion solutions} ---Here we consider the topologically stable excitations of the saturated state ${\bf n}=\hat{\vec{z}}$. First, we utilize the constraint ${\bf n}^2=1$ by expressing the 
direction of the magnetization in spherical coordinates, ${\bf n}=\sin\theta(\cos\phi\,\hat{\vec{x}}+\sin\phi\,\hat{\vec{y}})+\cos\theta\hat{\vec{z}}$. One can show \cite{Note3} that for the case $\mathscr{E}_\textsc{dmi}^{\textsc{n}}$ the total energy \eqref{eq:E-H} has a local minimum if $\phi=\chi$ and function $\theta=\theta(\rho)$ is determined by the equation
\begin{equation}\label{eq:theta-skyrm}
	\begin{split}
		(1+\epsilon\cos^2\theta)\nabla_\rho^2\theta-\sin\theta&\cos\theta\left(\frac{1+\epsilon}{\rho^2}+\epsilon\,\theta'^2\right)\\ 
		&+d\frac{\sin^2\theta}{\rho}-\sin\theta=0,
	\end{split}
\end{equation}
where we introduced the polar frame of reference $\{\rho,\chi\}$ with the radial distance $\rho$ measured in units of $\ell=\sqrt{2A/(M_sH)}$ and  $\nabla_\rho^2f=\rho^{-1}\partial_\rho(\rho\,\partial_\rho f)$ denotes radial part of the Laplace operator. Equation \eqref{eq:theta-skyrm} must be solved with the boundary conditions $\theta(0)=\pi$, $\theta(\infty)=0$. A number of examples of 
skyrmion profiles determined by Eq.~\eqref{eq:theta-skyrm} for various values of parameters $d$ and $\epsilon$ are shown in Fig.~S2 \cite{Note3}. Note that the skyrmion size is mainly determined by the parameter $d$, while the parameter $\epsilon$ weakly modifies the details of the skyrmion profile. For the case $\mathscr{E}_\textsc{dmi}^{\textsc{b}}$ the equilibrium solution is $\phi=\chi+\pi/2$ and the corresponding equation for the profile $\theta(\rho)$ coincides with \eqref{eq:theta-skyrm} when $\epsilon=0$. Note that in this case Eq.~\eqref{eq:theta-skyrm} is reduced to the well known skyrmion equation \cite{Leonov16,Bogdanov94,Bogdanov89}.
 
In order to analyze stability of the obtained static solutions we study spectrum of the skyrmion eigen-excitations by means of the methods commonly applied for skyrmions \cite{Kravchuk18,Schuette14} as well as for others two-dimensional magnetic topological solitons \cite{Ivanov98,Sheka01,Sheka04,Ivanov05b,Sheka06}.
Namely, we introduce time-dependent small deviations
$\theta=\theta_0+\vartheta$ and $\phi=\phi_0+\varphi/\sin\theta_0$, where $\vartheta,\,\varphi\ll1$ and $\theta_0=\theta_0(\rho)$, $\phi_0$ denotes the static profile. The linearization of the Landau-Lifshitz equations, $\sin\theta\partial_t\phi=\frac{\gamma}{M_s}\delta E_{\mathrm{eff}}/\delta\theta$, $-\sin\theta\partial_t\theta=\frac{\gamma}{M_s}\delta E_{\mathrm{eff}}/\delta\phi$, in the vicinity of the static solution results in solutions for the deviations in the form $\vartheta=f(\rho)\cos(\omega\tau+m\chi+\chi_0)$, $\varphi=g(\rho)\sin(\omega\tau+m\chi+\chi_0)$, where $m\in\mathbb{Z}$ is an azimuthal quantum number and$\chi_0$ ia an arbitrary phase.  Here $\tau=t\Omega_0$ is the dimensionless time, where  $\Omega_0=\gamma H$ is the Larmor frequency with $\gamma$ being the gyromagnetic ratio. The eigenfrequencies $\omega$ and the corresponding eigenfunctions $f$, $g$ are determined by 
solving the Bogoluybov-de Gennes eigenvalue problem \cite{Note3}. The numerical solution was obtained for a range of $d$ and a couple of values of $\epsilon$.  
A number of bounded eigenmodes with $\omega<1$ are found in the gap. Eigenfrequencies of the radially-symmetric ($m=0$) and elliptic ($m=2$) modes are shown in Fig.~\ref{fig:spectrum}, where we compare both types of DM terms \footnote{The stability analysis was performed for zero temperature. However, thermally induced magnons 
	would only result in additional damping for bounded skyrmion modes.}. If $\epsilon=0$, the spectra are identical for both cases, in particular, the well known elliptical instability \cite{Bogdanov94a,Schuette14} take place due to the softening of the elliptic mode in the region $d>d_c$, where the uniformly saturated state is thermodynamically unstable \cite{Schuette14}. For the case $\mathscr{E}_\textsc{dmi}^{\textsc{n}}$ the $\epsilon$-term shifts the elliptical instability to the larger values of $d$ with the condition $d>d_c^{\textsc{n}}(\epsilon)$ kept, while in the case $\mathscr{E}_\textsc{dmi}^{\textsc{b}}$ the effect of the $\epsilon$-term is negligible.

Remarkably, the $\epsilon$-term influences oppositely on the  breathing mode ($m=0$), for different DM types. For the case $\mathscr{E}_\textsc{dmi}^{\textsc{b}}$ the eigenfrequency $\omega_0$ is increased and for small $d$ the breathing mode is pushed out from the gap into the magnon continuum. As a result, the small-radius skyrmions are free of the bounded states. This is in contrast to the case $\mathscr{E}_\textsc{dmi}^{\textsc{n}}$, when the breathing mode eigenfrequency is rapidly decreased resulting in a radial instability for small $d$. In order to give some physical insight to the latter effect we consider the model, where the skyrmion profile is described by the linear Ansatz \cite{Bogdanov89,Bogdanov94} $\theta_{\text{a}}(\rho)=\frac{\pi}{R}(R-\rho)\mathrm{H}(R-\rho)$, and $\phi=\chi+\Phi$.
Here the variational parameters $R$ and $\Phi$ describe the skyrmion radius and helicity, respectively, and $\mathrm{H}(x)$ is the Heaviside step function. For this model total energy \eqref{eq:E-H} with $\mathscr{E}_\textsc{dmi}=\mathscr{E}_\textsc{dmi}^{\textsc{n}}$ reads
\begin{equation}\label{eq:E-ansatz}
\frac{E_\text{tot}^\textsc{n}}{2\pi AL}=e_{\text{ex}}+\epsilon e_\epsilon\cos^2\Phi-2\delta\cos\Phi R+R^2e_\textsc{h},
\end{equation}
where the constants $e_{\text{ex}}\approx6.15$ \footnote{The exact value is $e_\text{ex}=[\pi^2+\gamma_0-\mathrm{Ci}(2\pi)+\ln(2\pi)]/2$, where $\gamma_0$ is Euler constant and $\mathrm{Ci}(x)$ denotes the cosine integral function, see \cite{Bogdanov94}.}, $e_\epsilon=e_{\text{ex}}-\pi^2/4$ and $e_\textsc{h}=1-4/\pi^2$ originate from the exchange, $\epsilon$-term and Zeeman contributions, respectively. Here $\delta=d\,\pi/4$.
 The energy expression \eqref{eq:E-ansatz} shows that the equilibrium helicity $\Phi$ is determined by the competition of the $\epsilon$-term, which tends to $\Phi=\pm\pi/2$ (Bloch skyrmion), and the DM term, which tends to $\Phi=0$ (N\'eel skyrmion). In the same time, the equilibrium skyrmion radius is determined by the competiotion of the DM and Zeeman terms, and for the Bloch skyrmion one has $R=0$. Thus, the skyrmion collapse is expected with the $\epsilon$ increasing. Indeed,  the minimization of the total energy \eqref{eq:E-ansatz} with respect to the both variational parameters results in the critical value $\epsilon_c=\delta^2/(e_\epsilon e_\textsc{h})$: if $\epsilon<\epsilon_c$ then the equilibrium values of the variational parameters $R_0=\delta/e_\textsc{h}$ and $\Phi_0=0$ determines the N\'eel skyrmion; if $\epsilon>\epsilon_c$ that the minimum of energy \eqref{eq:E-ansatz} is reached for $R_0=0$ and $\Phi_0=\pm\pi/2$. The latter corresponds to a collapsed Bloch skyrmion.
In other words, a stable N\'eel slyrmion exists for the case $\epsilon<\epsilon_c$. Surprisingly, there are no intermediate states with $0<\Phi_0<\pi/2$ when $\epsilon>\epsilon_c$.

{\it Skyrmion lattice} --- In order to estimate the region of existence of the skyrmion lattice we use the circular cell approximation \cite{Bogdanov94}, when the lattice cell is approximated by a circle of radius $R$ and  the boundary
condition $\theta(R)=0$ is applied. The skyrmion profile is described by the same linear Ansatz as for the case of an isolated skyrmion. Minimizing the energy \eqref{eq:E-ansatz} per unit cell ${E}_\textsc{2d}^\textsc{n}=E_\mathrm{tot}^\textsc{n}/(\pi R^2)$ one obtains the following equilibrium values of the variational parameters $\Phi_0^\textsc{n}=0$, $R_0^\textsc{n}(\epsilon)=(e_{\text{ex}}+\epsilon e_\epsilon)/\delta$,  
and the corresponding equilibrium energy reads ${E}_\textsc{2d}^\textsc{n}(\epsilon)=2AL\left[e_\textsc{h}-\delta^2/(e_{\text{ex}}+\epsilon e_\epsilon)\right]$.
For the case $\mathscr{E}_\textsc{dmi}^{\textsc{b}}$ the same procedure results in the $\epsilon$-independent values: $\Phi_0^\textsc{b}=\pi/2$, $R_0^\textsc{b}=R_0^\textsc{n}(0)$ and ${E}_\textsc{2d}^\textsc{b}={E}_\textsc{2d}^\textsc{n}(0)$.

\begin{figure}
	\includegraphics[width=\columnwidth]{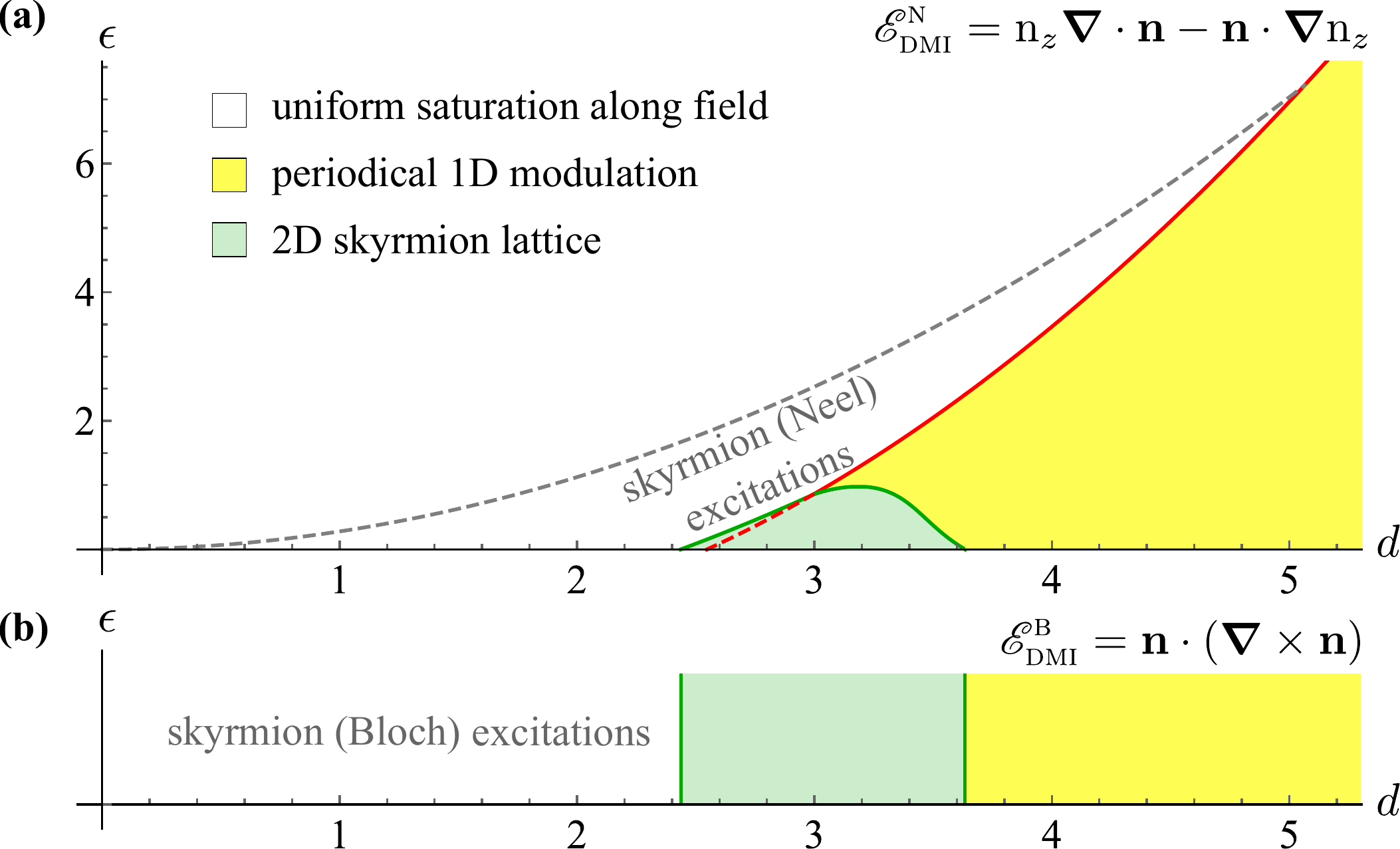}
	\caption{Diagrams of the ground states for different kinds of DMI. (a): the red line is determined by the condition $d=d_c^{\textsc{n}}(\epsilon)$, it separates the uniform state and periodical 1D modulation. The green region of the N\'eel skyrmion lattices is determined by the conditions ${E}_\textsc{2d}^\textsc{n}<{E}_\textsc{1d}^\textsc{n}$ and ${E}_\textsc{2d}^\textsc{n}<0$ to the right and to the left of the red line. The gray dashed line is the line of collapse of the N\'eel skyrmions, it is determined by the condition $\epsilon=\epsilon_c(d)$. (b): colors have the same meaning as on the panel (a), but periodical helical state and skyrmion lattices are of Bloch type. The excitations in form of isolated Bloch skyrmions are stable within all white region. If ${\bf d}_1$ is used in the Hamiltonian (1), the coefficient of the $(\nabla\cdot{\bf n})^2$ would be negative, which would in turn make the Bloch skyrmion more stable}\label{fig:diagram}
\end{figure}

Comparing energies of three states, namely, the energy of the uniform magnetization along field $E=0$, energy of the 1D periodical state (per period) $E_{\textsc{1d}}$, and energy of the skyrmion lattice per unit cell $E_{\textsc{2d}}$, we determine the phase diagram of the ground states, see Fig.~\ref{fig:diagram}.
Note that for $\epsilon>\epsilon_0\approx0.98$ the skyrmion lattice is not a ground state. Given the dependence of $\epsilon$ with the 
exchange coupling $J_0$, temperature, and chemical potential, the skyrmion lattice phase is likely to occur for a not 
too high temperature range as compared to the Curie temperature of EuS. . 
The dimensionless DM parameter $d$ can then be tuned by the external field to attain the 
interval shown under the green area of Fig. \ref{fig:diagram}(a). 

{\it  Conclusions} --- We have shown that the effective magnetic energy for a TI-FMI heterostructure exhibits a Dzyaloshinskii-Moriya term induced 
by tracing out the surface Dirac fermions proximate to the FMI. A unique feature of the effective energy as compared to other 
DM systems is the presence of an additionally induced magnetic capacitance energy, given by a term proportional to the square of the 
magnetic charge $\nablab\cdot{\bf n}$. Despite having a small magnitude in realistic samples, the interplay between this term and 
the DM one yields a phase diagram with interesting phase boundaries in the case of a N\'eel DMI, which is the situation relevant 
for, e.g., Bi$_2$Se$_3$ samples proximate to a FMI. 
Our theory is directly relevant for very recently synthesized  TI - ferromagnetic thin film heterostructures, in some of which the formation of a skyrmionic magnetic texture has been observed~\cite{Zhang18}. 

\acknowledgments

F.S.N. and I.E. thank the DFG Priority Program SPP 1666, "Topological Insulators", under Grant number, ER 463/9. 
J.v.d.B. acknowledges support from SFB 1143.  FK and JSM. acknowledge the support from NSF Grants No. DMR-1207469, 1700137 ONR Grant No. N00014-13-1- 0301, N00014-16-1-2657 and the STC Center for Integrated Quantum Materials under NSF Grant No. DMR-1231319. 
F.K. acknowledges the Science and Technological Research Council of Turkey (TUBITAK) through the BIDEB 2232 Program under award number 117C050 (Low--Dimensional Hybrid Topological Materials). I.E. acknowledges support by the Ministry of Education and Science of the Russian Federation in the framework of
Increase Competitiveness Program of NUST MISiS (K2-2017-085).   

\bibliography{div-term}

\appendix
	
	\section{Supplemental Information}
	\renewcommand\theequation{S\arabic{equation}}
	\setcounter{equation}{0}
	\renewcommand\thefigure{S\arabic{figure}}
	\setcounter{figure}{0}
	
	\subsection{Effective free energy and induced DM term}
	
	Let us consider the following low energy Hamiltonian for the Dirac fermions on the surface of the TI, 
	\begin{equation}
	\label{Eq:Dirac}
	H_{\rm Dirac}=\left[{\bf d}\left(-i\hbar\nablab\right)-J_0{\bf n}({\bf r})\right]\cdot\sigmab.
	\end{equation}
	
	Quite generally, the non-interacting Green function for a spin-momentum locked system can be written, 
	\begin{equation}
	{\cal G}_{\alpha\beta}(X)=G(X)\delta_{\alpha\beta}+{\bf F}(X)\cdot\sigmab_{\alpha\beta},
	\end{equation}
	where $X=(\tau,{\bf r})$, with $\tau\in[0,\beta]$ being the imaginary time.
	Thus, expanding the free energy expression up to $J_0^2$, we obtain 
	${F}_{\rm Dirac}={F}_{\rm Dirac}(0)+\delta{F}_{\rm Dirac}$, where,    
	\begin{widetext}
		\begin{equation}
		\delta{F}_{\rm Dirac}=\frac{1}{2}\int_X\int_{X'}{\cal G}_{\alpha\beta}(X-X'){\cal G}_{\gamma\delta}(X'-X)
		[J_0{\bf n}(X)\cdot\sigmab_{\beta\gamma}]
		[J_0{\bf n}(X')\cdot\sigmab_{\delta\alpha}], 
		\end{equation}
		where we use the shorthand notation $\int_X\equiv\int_0^\beta d\tau\int d^2r$. Lengthy but straightforward calculations  yield, 
		\begin{eqnarray}
		\label{Eq:mag}
		\delta{F}_{\rm Dirac}^{\rm mag}&=&\frac{J_0^2}{2}\int_X\int_{X'}\{2[G(X-X')G(X'-X)-{\bf F}(X-X')\cdot{\bf F}(X'-X)]{\bf n}(X)\cdot{\bf n}(X')
		\nonumber\\
		&+&4{\bf F}(X-X')\cdot{\bf n}(X){\bf F}(X'-X)\cdot{\bf n}(X')-24iG(X-X'){\bf F}(X'-X)\cdot[{\bf n}(X)\times{\bf n}(X')]\},
		\end{eqnarray}  
	\end{widetext}
	where in writing the above equation we have made use of the property ${\bf F}(\tau,-{\bf r})=-{\bf F}(\tau,{\bf r})$. 
	
	Observe that Eq. (\ref{Eq:mag}) features also the general Dzyloshinsky-Moriya (DM) term, which arises in the third line of that equation. It can be re-cast in a more familiar form by specializing 
	to the simple case, 
	\begin{equation}
	G(\omega_n,{\bf k})=\frac{i\omega_n+\mu}{(i\omega_n+\mu)^2-{\bf d}^2({\bf k})},
	\end{equation}
	\begin{equation}
	{\bf F}(\omega_n,{\bf k})=-\frac{{\bf d}({\bf k})}{(i\omega_n+\mu)^2-{\bf d}^2({\bf k})}, 
	\end{equation}
	where $\omega_n=(2n+1)\pi/\beta$ and ${\bf d}({\bf k})$ is either ${\bf d}_1$ or ${\bf d}_2$ from Eq. (2) in momentum space.  
	Assuming a time-independent magnetization, Eq. (\ref{Eq:mag}) becomes,  
	\begin{eqnarray}
	\label{Eq:dF-1}
	\delta{F}_{\rm Dirac}^{\rm mag}=\frac{J_0^2}{2}\int\frac{d^2k}{(2\pi)^2}
	[S_{ab}({\bf k})+A_c({\bf k})\epsilon_{abc}]n_a({\bf k})n_b(-{\bf k}),
	\nonumber\\
	\end{eqnarray}
	featuring symmetric and antisymmetric contributions $S_{ab}$ and $A_c$, which are given by, 
	\begin{widetext}
		\begin{equation}
		\label{Eq:sym}
		S_{ab}({\bf k})=\frac{2}{\beta}\sum_{n=-\infty}^{\infty}\int\frac{d^2q}{(2\pi)^2}
		\frac{[(i\omega_n+\mu)^2-{\bf d}({\bf q})\cdot{\bf d}({\bf q}+{\bf k})]\delta_{ab}+2d_{a}({\bf q})d_{b}({\bf q}+{\bf k})}
		{[(i\omega_n+\mu)^2-(\hbar v_F)^2q^2][(i\omega_n+\mu)^2-(\hbar v_F)^2({\bf q}+{\bf k})^2]},
		\end{equation}
		\begin{equation}
		\label{Eq:antisym}
		A_c({\bf k})=\frac{24i}{\beta}\sum_{n=-\infty}^{\infty}\int\frac{d^2q}{(2\pi)^2}\frac{(i\omega_n+\mu)d_c({\bf q})}
		{[(i\omega_n+\mu)^2-(\hbar v_F)^2q^2][(i\omega_n+\mu)^2-(\hbar v_F)^2({\bf q}+{\bf k})^2]},
		\end{equation}
	\end{widetext}
	respectively.  
	
	The symmetric contribution induces a magnetization stiffness at the interface. 
	Part of the integral yielding $S_{ab}$ does not converge and needs to be regularized by means of  
	a cutoff $\Lambda$. The result in the long wavelength limit is, 
	\begin{equation}
	\label{Eq:Sab}
	S_{ab}({\bf k})=-\frac{\Lambda}{4\pi\hbar v_F}\delta_{ab}+{\cal S}(\beta, \mu)k^2\left(\delta_{ab}+\frac{k_ak_b}{k^2}\right),
	\end{equation} 
	where ${\cal S}(\beta,\mu)=\beta/[24\pi\cosh^2(\beta\mu/2)]$.
	
	The antisymmetric contribution vanishes for $\mu=0$, since the Matsubara sum involves an odd summand in this case. 
	Thus, in order to generate a DM term the chemical potential must be nonzero. Assuming $\mu\neq 0$, 
	and a long wavelength limit, we obtain,   
	\begin{eqnarray}
	A_c({\bf k})=-id_c({\bf k}){\cal A}(\beta\mu),
	\end{eqnarray}
	where ${\cal A}(\beta\mu)=(8\pi\hbar v_F)^{-1}\tanh\left(\beta\mu/2\right)$.
	Writing, 
	\begin{equation}
	F_{\rm Dirac}^{\rm mag}=\int d^2r {\cal F}_{\rm Dirac}^{\rm mag}({\bf n}({\bf r})),
	\end{equation} 
	we obtain for the fluctuation correction of the  free energy density,
	\begin{eqnarray}
	\label{Eq:FE-Dirac}
	\delta{\cal F}_{\rm Dirac}^{\rm mag}&=&\frac{s}{2}\left\{[\nablab{\bf n}({\bf r})]^2 
	+[\nablab\cdot{\bf n}({\bf r})]^2\right\}
	\nonumber\\
	&+&i\frac{a}{2} {\bf n}({\bf r})\cdot[{\bf d}(-i\hbar\nablab)\times{\bf n}({\bf r})],
	\end{eqnarray}
	where $(\nablab{\bf n})^2=\sum_{i=x,y,z}(\nablab n_i)^2$ defines the usual exchange term, and 
	we have defined $s=J_0^2{\cal S}(\beta,\mu)$ and $a=J_0^2{\cal A}(\beta,\mu)$. 
	A term proportional to $\Lambda{\bf n}^2$ implied by the first term in (\ref{Eq:Sab}) has been removed, since it is actually 
	a constant in view of the constraint ${\bf n}^2=1$. Similarly, we can drop the constant term 
	$F_{\rm Dirac}(0)$ from the free energy, since it does not depend on the field. Thus, we can safely write 
	${\cal F}_{\rm Dirac}=\delta{\cal F}_{\rm Dirac}$. 
	The above expression features a DM term induced by Dirac fermion fluctuations. In addition, a contribution $\sim (\nablab\cdot{\bf n})^2$ 
	is also generated. This term leads to interesting physical properties when ${\bf d}_2$ is replaced for 
	${\bf d}$ in Eq. (\ref{Eq:FE-Dirac}), modifying in this way the behavior of N\'eel skyrmions.  
	Note that differently from the case where the Dirac fermion is gapped \cite{Rex16a}, no intrinsic anisotropy is generated by the Dirac fermions. 
	
	\subsection{One-dimensional magnetization modulation}
	Here we consider one-dimensional static solutions ${\bf n}={\bf n}(x)$ of the model (8). Let us start with the case $\mathscr{E}_\textsc{dmi}=\mathscr{E}_\textsc{dmi}^{\textsc{n}}$.
	In this case the effective energy $E_{\rm eff}=AL\int\mathcal{E}_{\textsc{1d}}^{\textsc{n}}\mathrm{d}^2x$ is described by  the density
	\begin{equation}\label{eq:E-den-1D}
	\begin{split}
	\mathcal{E}_{\textsc{1d}}^{\textsc{n}}&=\theta'^2+\sin^2\theta\phi'^2+\epsilon\left(\cos\theta\cos\phi\theta'-\sin\theta\sin\phi\phi'\right)^2\\
	&+d\left(\cos\phi\theta'-\sin\theta\cos\theta\sin\phi\phi'\right)+2(1-\cos\theta),
	\end{split}
	\end{equation}
	where angles $\theta(x)$ and $\phi(x)$ determine orientation of the unit magnetization vector  ${\bf n}=\sin\theta(\cos\phi\,\hat{\vec{x}}+\sin\phi\,\hat{\vec{y}})+\cos\theta\hat{\vec{z}}$, prime denotes the derivation with respect to the dimensionless coordinate $x$ measured in units of $\ell$. The corresponding Euler-Lagrange equations $\delta E_{\rm eff}/\delta\theta=0$ and $\delta E_{\rm eff}/\delta\phi=0$ are as follows
	\begin{subequations}\label{eq:theta-phi-1D}
		\begin{align}
		\label{eq:theta-1D}	&\theta''-\sin\theta\cos\theta\phi'^2-\epsilon\cos\theta\cos\phi\,\Xi\\ \nonumber
		&-d\sin^2\theta\sin\phi\phi'-\sin\theta=0,\\
		\label{eq:phi-1D}	&\left(\sin^2\theta\phi'\right)'+\epsilon\sin\theta\sin\phi\,\Xi+d\sin^2\theta\sin\phi\theta'=0,
		\end{align}
		where
		\begin{align}
		\Xi&=\cos\theta\left[2\sin\phi\,\theta'\phi'-\cos\phi\,\theta''\right]\\ \nonumber &+\sin\theta\left[\cos\phi\left(\theta'^2+\phi'^2\right)+\sin\phi\,\phi''\right].
		\end{align}	
	\end{subequations} 
	First of all, Eqs.~\eqref{eq:theta-phi-1D} have trivial solution $\theta=0$, which corresponds to the uniform magnetization along the field ${\bf n}=\hat{\vec{z}}$. The uniform state has zero energy $\mathcal{E}_{\textsc{1d}}^{\textsc{n}}=0$. The another trivial solution $\theta=\pi$ (${\bf n}=-\hat{\vec{z}}$) corresponds to the energy maximum and it is unstable.
	
	Let us consider possible nonuniform solutions. Equation \eqref{eq:phi-1D} is satisfied when $\sin\phi=0$ ($\phi=0$ for $d\theta'<0$, and $\phi=\pi$ for $d\theta'>0$), or in the other words $\mathrm{n}_y=0$. Thus, the magnetization lies within the plane $\hat{\vec{x}}0\hat{\vec{z}}$. The components $\mathrm{n}_x=\sin\theta$ and $\mathrm{n}_z=\cos\theta$ are determined by the equation \eqref{eq:theta-1D}, which now looks has a form
	\begin{equation}\label{eq:theta-period}
	\theta''\left(1+\epsilon\cos^2\theta\right)=\sin\theta\left(1+\epsilon\cos\theta\theta'^2\right),
	\end{equation}
	and the corresponding energy density reads
	\begin{equation}\label{eq:E-1D}
	\mathcal{E}=\theta'^2(1+\epsilon\cos^2\theta)+4\sin^2\frac{\theta}{2}-d\theta',
	\end{equation}
	where we assumed that $d\theta'>0$.
	Rewriting the Eq.~\eqref{eq:theta-period} in form $\left[\theta'^2(1+\epsilon\cos^2\theta)\right]'=-2(\cos\theta)'$ one can easily find its first integral
	\begin{equation}\label{eq:theta-first-int}
	\theta'^2(1+\epsilon\cos^2\theta)=4\sin^2\frac{\theta}{2}+4C,
	\end{equation}
	where $C$ is the integration constant. Equation \eqref{eq:theta-first-int} admits separation of the variables and can be solved in quadratures. The constant $C$ determines period $T$ of the magnetization structure: ${\bf n}(x+T)={\bf n}(x)$. Let us first consider the particular case $C=0$, it corresponds to a solution of \eqref{eq:theta-first-int}, which satisfies the boundary conditions $\theta(-\infty)=0$, $\theta(+\infty)=2\pi$ (or vice-versa), this is a single $2\pi$-domain wall of N\'eel type, in this case $T\to\infty$. Taking into account \eqref{eq:theta-first-int} and \eqref{eq:E-1D} one can present energy ${E}_\textsc{dw}=\int_{-\infty}^{+\infty}\mathcal{E}\text{d}x$ of this domain wall in form
	\begin{equation}\label{eq:E-DW}
	E_\textsc{dw}=8\int\limits_0^\pi\sin\theta\sqrt{1+\epsilon\cos^22\theta}\,\text{d}\theta-2\pi d.
	\end{equation}
	The condition $E_\textsc{dw}<0$, or equivalently
	\begin{equation}\label{eq:dc-cond}
	d>d_c^{\textsc{n}}(\epsilon)=\frac{8}{\pi}\int\limits_{0}^{1}\sqrt{1+\epsilon(2\xi^2-1)^2}\mathrm{d}\xi,
	\end{equation}
	determines the area of parameters, where nucleation of the $2\pi$-domain wall on the background of the saturated state ${\bf n}=\hat{\vec{z}}$ is energetically preferable. 
	
\begin{figure}
		\includegraphics[width=\columnwidth]{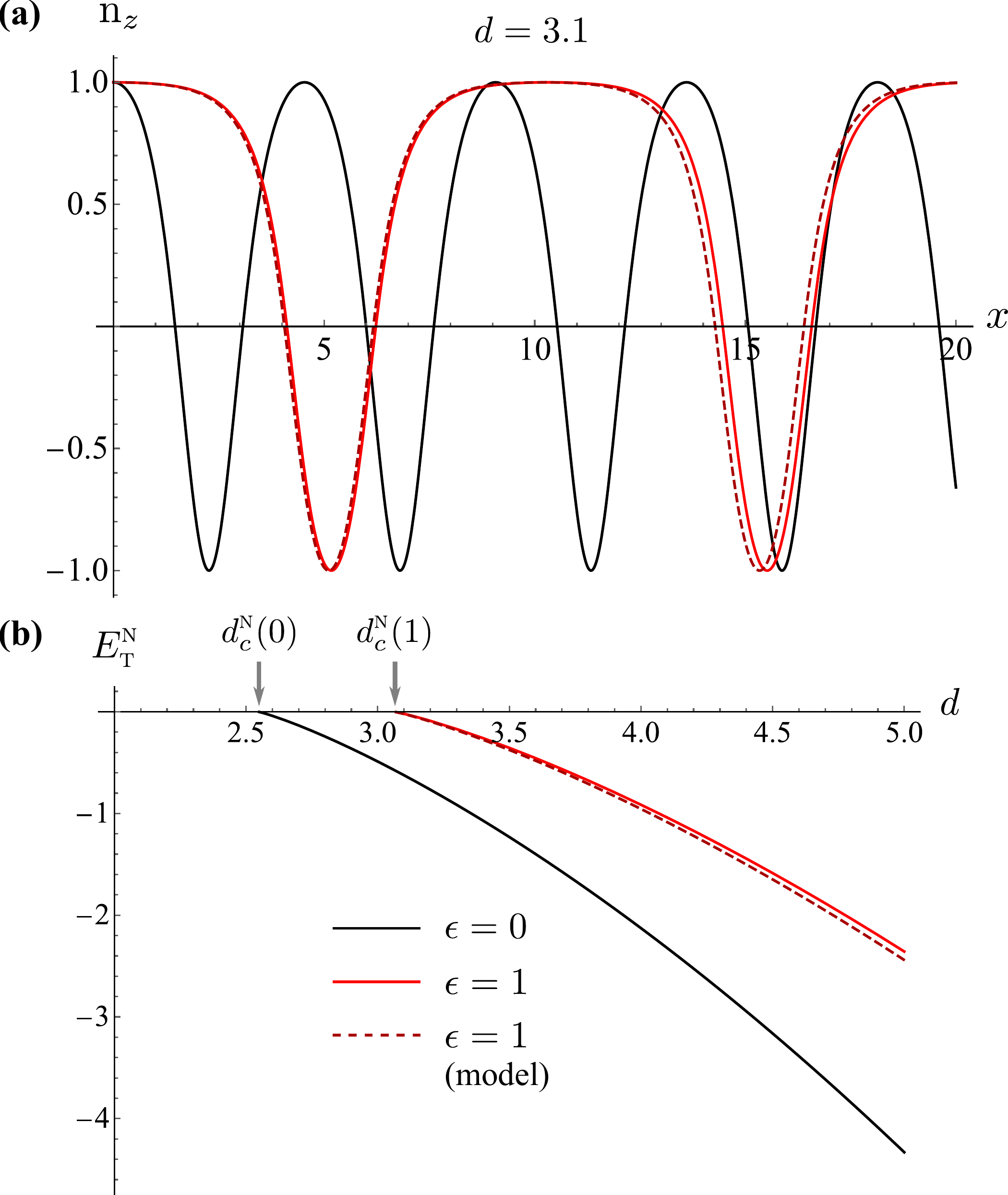}
		\caption{Influence of the $\epsilon$-term on 1D periodical structure. (a) -- perpendicular magnetization component $\mathrm{n}_z$; (b) -- energy per period. The dashed line corresponds to the approximations: (a) the solution of Eq.~\eqref{eq:theta-first-int} with $C=\tilde{C}$, where $\tilde{C}$ is determined by \eqref{eq:C-tilde}; (b) $\tilde{E}^\textsc{n}_\textsc{t}=-4\tilde{C}$. }\label{fig:1D-modulation}
	\end{figure}

	The general case $C>0$ corresponds to the helical state, which can be interpreted as a periodical sequence of the considered domain walls. If $\epsilon=0$ \footnote{The case $\epsilon=0$ was analysed earlier in Ref.~\cite{Bogdanov94}.}, then the solution of \eqref{eq:theta-first-int} reads
	\begin{equation}\label{eq:theta-am}
	\theta=2\,\mathrm{am}\left(\sqrt{C}(x-x_0)\left|-\frac{1}{C}\right.\right),
	\end{equation}
	where $\mathrm{am}(x|k)$ is Jacobi's amplitude function \cite{Abramowitz72} and the integration constant $x_0$ determines a uniform shift along $x$-axis. 
	The solution \eqref{eq:theta-am} determines period of the magnetization components $\mathrm{n}_{x}=\sin\theta$ and $\mathrm{n}_z=\cos\theta$:
	\begin{equation}\label{eq:T}
	T=\frac{2}{\sqrt{C}}\mathrm{K}\left(-\frac{1}{C}\right),
	\end{equation}
	where $\mathrm{K}(k)$ is complete elliptic integral of the first kind \cite{Abramowitz72}. The constant $C$ must be found from the minimization of the total energy per period $E_{\textsc{t}}^{\textsc{n}}=T^{-1}\int_{0}^{T}\mathcal{E}\,\mathrm{d}x$. Performing the minimization procedure for \eqref{eq:theta-am} and \eqref{eq:T} one obtains $E_{\textsc{t}}^{\textsc{n}}=-4C$ for energy of the equilibrium structure, where the equilibrium value of the constant $C$ is determined by the equation
	\begin{equation}\label{eq:C}
	\frac{d}{d_c^{\textsc{n}}(0)}=\sqrt{C}\,\mathrm{E}\left(-\frac{1}{C}\right)
	\end{equation}
	with  $\mathrm{E}(k)$ being the complete elliptic integral of the second kind. The total energy (8) per period reads $E_{\textsc{1d}}^{\textsc{n}}=ALE_{\textsc{t}}^{\textsc{n}}$. An example of a solution for the case $\epsilon=0$ is shown in Fig.~\ref{fig:1D-modulation}(a) by the black line.
	
	For the case $\epsilon>0$, the solution of Eq.~\eqref{eq:theta-first-int}, which minimizes the energy $E_{\textsc{t}}^{\textsc{n}}(C)$ with respect to $C$ can be found only numerically. An example for the case $\epsilon=1$ is shown in Fig.~\ref{fig:1D-modulation}(a) by the red solid line. As one can see, the $\epsilon$-term can noticeably increase period of the structure. It is important to note that one can avoid the tedious procedure of the numerical minimization of the energy $E_{\textsc{t}}^{\textsc{n}}(C)$ by using the fact that the equilibrium value $C\approx\tilde{C}$, where $\tilde{C}$ is found in the way analogous to \eqref{eq:C}:
	\begin{equation}\label{eq:C-tilde}
	\frac{d}{d_c(\epsilon)}=\sqrt{\tilde C}\,\mathrm{E}\left(-\frac{1}{\tilde C}\right).
	\end{equation}
	An example of such an approximate solution is shown in Fig.~\ref{fig:1D-modulation}(a) by the red dashed line. And the corresponding energy per period can be well approximated as $E_{\textsc{t}}^{\textsc{n}}\approx-4\tilde C$, see Fig.~\ref{fig:1D-modulation}(b). 
	
	Let us now proceed to the case $\mathscr{E}_\textsc{dmi}=\mathscr{E}_\textsc{dmi}^{\textsc{b}}$.  In this case the energy density coincides with \eqref{eq:E-den-1D} but the DMI-term
	\begin{equation}\label{eq:E-den-1D-B}
	\begin{split}
	\mathcal{E}_{\textsc{1d}}^{\textsc{b}}=&\theta'^2+\sin^2\theta\phi'^2+\epsilon\left(\cos\theta\cos\phi\theta'-\sin\theta\sin\phi\phi'\right)^2\\
	&+d\left(\sin\phi\theta'+\sin\theta\cos\theta\cos\phi\phi'\right)+2(1-\cos\theta).
	\end{split}
	\end{equation}
	Energy expression \eqref{eq:E-den-1D-B} generates the Euler-Lagrange equations
	\begin{subequations}\label{eq:theta-phi-1D-B}
		\begin{align}
		\label{eq:theta-1D-B}	\theta''&-\sin\theta\cos\theta\phi'^2-\epsilon\cos\theta\cos\phi\,\Xi\\ \nonumber
		&+d\sin^2\theta\cos\phi\phi'-\sin\theta=0,\\
		\label{eq:phi-1D-B}	\left(\sin^2\theta\phi'\right)'&+\epsilon\sin\theta\sin\phi\Xi-d\sin^2\theta\cos\phi\theta'=0,
		\end{align}
	\end{subequations}
	which coincide with \eqref{eq:theta-phi-1D} up to the replacement $\sin\phi\leftrightarrow-\cos\phi$ in the DMI terms. As a result, equation \eqref{eq:phi-1D-B} is satisfied when $\cos\phi=0$ (in this case $\Xi=0$), or in the other words $\mathrm{n}_x=0$. Thus, the magnetization lies within the plane $\hat{\vec{y}}0\hat{\vec{z}}$. The latter corresponds to the Bloch domain walls. The components $\mathrm{n}_y=\sin\theta$ and $\mathrm{n}_z=\cos\theta$ are determined by the equation \eqref{eq:theta-1D-B}, which coincides with \eqref{eq:theta-period} if $\epsilon=0$. Thus, $\epsilon$ does not effect the static 1D solution for the case $\mathscr{E}_\textsc{dmi}=\mathscr{E}_\textsc{dmi}^{\textsc{b}}$ and the further analysis coincides with the one done for $\mathscr{E}_\textsc{dmi}=\mathscr{E}_\textsc{dmi}^{\textsc{n}}$ with $\epsilon=0$.
	
	\subsection{Skyrmion solutions}
	
\begin{figure}
		\includegraphics[width=\columnwidth]{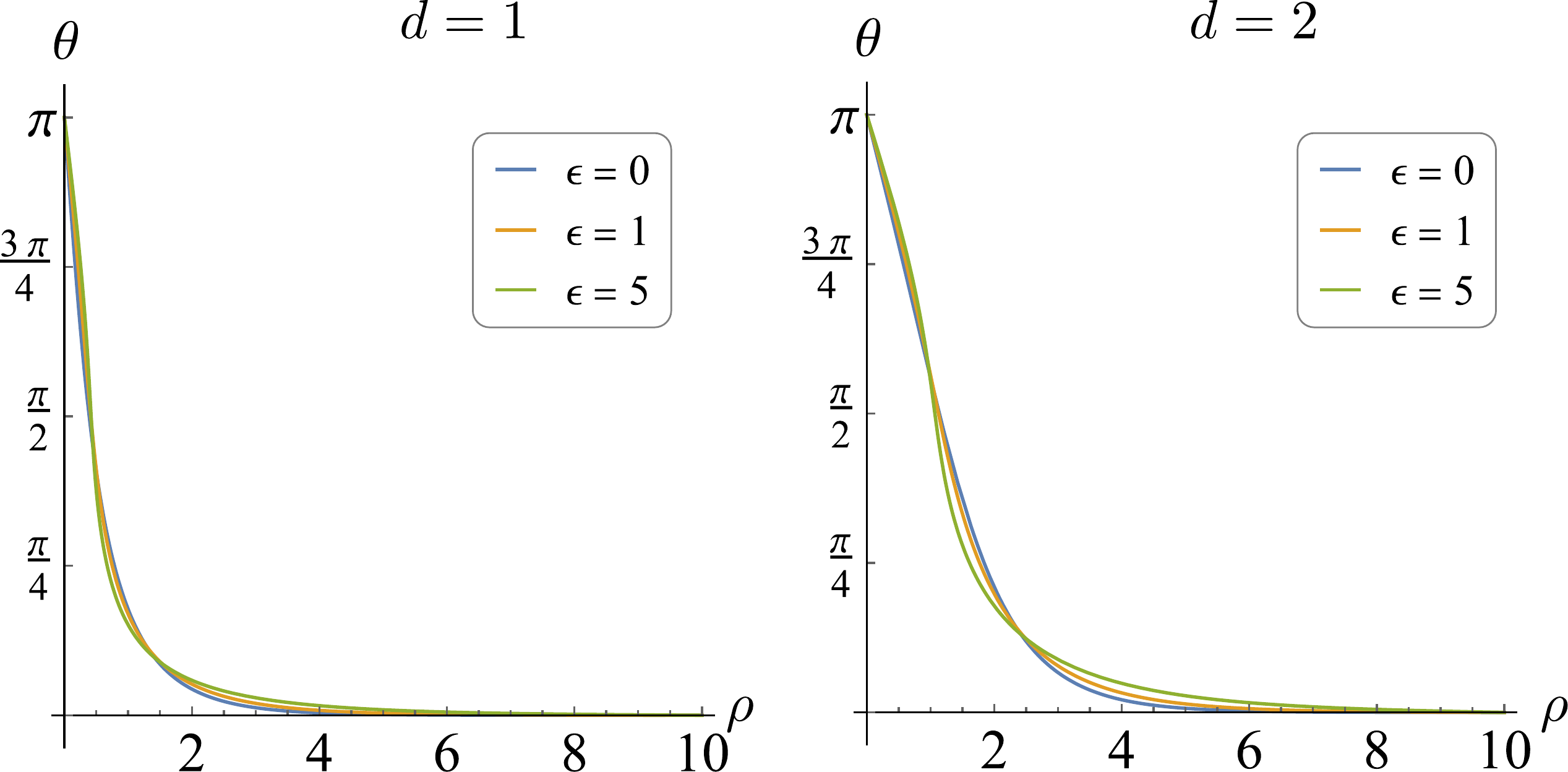}
		\caption{Skyrmion profiles determined by Eq.~(9) for various values of parameters $d$ and $\epsilon$.}\label{fig:neel-skrms-fld}
	\end{figure}

	Let us consider two-dimensional solutions. As previously, we first consider the case $\mathscr{E}_\textsc{dmi}=\mathscr{E}_\textsc{dmi}^{\textsc{n}}$.
	Introducing the normalized in-plane magnetization component $\vec{\eta}=\cos\phi\hat{\vec{x}}+\sin\phi\hat{\vec{y}}$ one can present the energy density in the form
	\begin{equation}\label{eq:E-2D}
	\begin{split}
	\mathcal{E}_{\textsc{2d}}^{\textsc{n}}=&(\vec{\nabla}\theta)^2+\sin^2\theta(\vec{\nabla}\phi)^2+\epsilon\left[\vec{\nabla}\cdot\left(\sin\theta\vec{\eta}\right)\right]^2\\
	&+2d\,\sin^2\theta\left(\vec{\eta}\cdot\vec{\nabla}\theta\right)+4\sin^2\frac{\theta}{2}.
	\end{split}
	\end{equation}
	The corresponding Euler-Lagrange equations $\delta E_{\rm eff}/\delta\theta=0$ and $\delta E_{\rm eff}/\delta\phi=0$  read
	\begin{subequations}\label{eq:theta-phi-2D}
		\begin{align}
		\label{eq:theta-2D}	&\nabla^2\theta+\epsilon\cos^2\theta\,\vec{\nabla}\cdot\left[\vec{\eta}\left(\vec{\nabla}\theta\cdot\vec{\eta}\right)\right]\\ \nonumber
		&-\sin\theta\cos\theta\left\{(\vec{\nabla}\phi)^2+\epsilon\left[(\vec{\nabla}\theta\cdot\vec{\eta})^2-\vec{\eta}\cdot\vec{\nabla}(\vec{\nabla}\cdot\vec{\eta})\right]\right\}\\ \nonumber
		&+d\sin^2\theta\vec{\nabla}\cdot\vec{\eta}-\sin\theta=0,\\
		\label{eq:phi-2D}&\vec{\nabla}\cdot\left[\sin^2\theta\vec{\nabla}\phi\right]-d\sin^2\theta(\bar{\vec{\eta}}\cdot\vec{\nabla}\theta)\\ \nonumber
		&+\epsilon\sin^2\theta\left[\bar{\vec{\eta}}\cdot\vec{\nabla}(\vec{\nabla}\cdot\vec\eta)-(\vec{\nabla}\theta\cdot\vec{\eta})(\vec{\nabla}\theta\cdot\bar{\vec{\eta}})\right]\\ \nonumber
		&+\epsilon\sin\theta\cos\theta\left[(\vec{\nabla}\theta\cdot\bar{\vec{\eta}})\vec{\nabla}\cdot\vec\eta+\bar{\vec{\eta}}\cdot\vec{\nabla}(\vec{\nabla}\theta\cdot\vec{\eta})\right]=0,
		\end{align}
	\end{subequations}
	where $\bar{\vec{\eta}}=\partial_\phi\vec{\eta}=-\sin\phi\hat{\vec{x}}+\cos\phi\hat{\vec{y}}$ is the unit vector perpendicular to $\vec{\eta}$. Equations \eqref{eq:theta-phi-2D} have trivial solution $\theta=0$. Besides, the Eq.~\eqref{eq:phi-2D} is always satisfied if $\vec{\eta}=\textbf{const}$ and $\theta=\theta(\xi)$ with $\xi$ being a coordinate along $\vec{\eta}$. This is the one-dimensional case considered in the previous section. The analogous ``one-dimensional'' solution takes place in the curvilinear polar frame of reference $\{\rho,\chi\}$. Indeed, in this case the Eq.~\eqref{eq:phi-2D} is satisfied if $\vec{\eta}=\vec{e}_\rho$ (equivalently $\phi=\chi$) and $\theta=\theta(\rho)$. Herewith, Eq.~\eqref{eq:theta-2D} is reduced to Eq.~(9),
	which describes profile of the isolated skyrmion. Note that for the case $d<0$ the in-plane magnetization is $\vec{\eta}=-\vec{e}_\rho$ (equivalently $\phi=\chi+\pi$). Few examples of skyrmion profiles determined by Eq.~(9) for various values of parameters $d$ and $\epsilon$ are shown in Fig.~\ref{fig:neel-skrms-fld}. Note that the skyrmion size is mainly determined by the parameter $d$, while the parameter $\epsilon$ weakly modifies details of the skyrmion profile.
	
	Let us now consider stability of the static skyrmion solutions of Eq. (9). To this end we introduce small deviations $\theta=\theta_0+\vartheta$ and $\phi=\phi_0+\varphi/\sin\theta_0$ of the static profile $\theta_0=\theta_0(\rho)$, $\phi_0=\chi$. Landau-Lifshitz equations $\sin\theta\partial_t\phi=\frac{\gamma}{M_s}\delta E_{\mathrm{eff}}/\delta\theta$, $-\sin\theta\partial_t\theta=\frac{\gamma}{M_s}\delta E_{\mathrm{eff}}/\delta\phi$ linearized in vicinity of the static solution with respect to the deviations $\vartheta$ and $\varphi$ are as follows
	\begin{equation}\label{eq:lin}
	\begin{aligned}
	\dot{\varphi}&=\hat{\mathrm{H}}_1\vartheta+W_1\partial_\chi\varphi+V\partial_{\rho\chi}^2\varphi,\\
	-\dot{\vartheta}&=\hat{\mathrm{H}}_2\varphi-W_2\partial_\chi\vartheta+V\partial_{\rho\chi}^2\vartheta,
	\end{aligned}
	\end{equation}
	where dot indicates the derivative with respect to the dimensionless time $\tau=t\Omega_0$ with $\Omega_0=\gamma H$ being the Larmor frequency. The differential operators and the potentials read
	\begin{equation}\label{eq:potentials}
	\begin{aligned}
	&\hat{\mathrm{H}}_1=-\nabla_\rho^2-\frac{\epsilon}{\rho}\partial_\rho\left(\rho\cos^2\theta_0\partial_\rho\right)-\frac{1}{\rho^2}\partial_{\chi\chi}^2+U_1,\\
	&\hat{\mathrm{H}}_2=-\nabla_\rho^2-\frac{1+\epsilon}{\rho^2}\partial_{\chi\chi}^2+U_2,\\
	&U_1=\cos2\theta_0\left(\frac{1+\epsilon}{\rho^2}+\epsilon\,\theta_0'^2\right)+\epsilon\sin2\theta_0\nabla_\rho^2\theta_0\\
	&-\frac{d}{\rho}\sin2\theta_0+\cos\theta_0,\\ &U_2=(1+\epsilon)(\cot\theta_0\nabla_\rho^2\theta_0-\theta_0'^2)-\frac{\epsilon}{\rho^2}-d\theta_0',\\
	&W_1=\frac{2+\epsilon}{\rho^2}\cos\theta_0-\frac{d}{\rho}\sin\theta_0,\\ &W_2=W_1-\frac{\epsilon}{\rho}\sin\theta_0\theta_0',\qquad V=-\frac{\epsilon}{\rho}\cos\theta_0.
	\end{aligned}
	\end{equation} 
	Note that term $(\nablab\cdot{\bf n})^2$ leads to the mixing of the derivatives in the linearized equations \eqref{eq:lin} (the $V$-term).  This is in contrast to the corresponding linear equations previously obtained for magnons over precessional solitons in easy-axis magnets \cite{Sheka01,Ivanov05b,Sheka06}, magnetic vortices in easy-plane magnets \cite{Ivanov98,Sheka04}, and magnetic skyrmions \cite{Schuette14,Kravchuk18}. 
	
	Equations \eqref{eq:lin} have solution $\vartheta=f(\rho)\cos(\omega\tau+\mu\chi+\chi_0)$, $\varphi=g(\rho)\sin(\omega\tau+\mu\chi+\chi_0)$, where $\mu\in\mathbb{Z}$ is azimuthal quantum number and $\chi_0$ ia an arbitrary phase. The eigenfrequencies $\omega$ and the corresponding eigenfunctions $f$, $g$ are determined by the following generalized eigen-value problem (EVP)
	\begin{equation}\label{eq:EVP}
	\hat{\mathcal{H}}\vec{\psi}=\omega\hat{\sigma}_x\vec{\psi},\quad \hat{\mathcal{H}}=\begin{Vmatrix}
	\hat{\mathscr{H}}_1&\mu(W_1+V\partial_\rho)\\
	\mu(W_2-V\partial_\rho)&\hat{\mathscr{H}}_2
	\end{Vmatrix}
	\end{equation}
	where $\vec{\psi}=(f,g)^\textsc{t}$ and $\hat{\sigma}_x$ is the first Pauli matrix. The diagonal operators are as follows
	\begin{equation}\label{eq:H}
	\begin{split}
	&\hat{\mathscr{H}}_1=-\nabla_\rho^2-\frac{\epsilon}{\rho}\partial_\rho\left(\rho\cos^2\theta_0\partial_\rho\right)+\frac{\mu^2}{\rho^2}+U_1\\
	&\hat{\mathscr{H}}_2=-\nabla_\rho^2+\frac{(1+\epsilon)\mu^2}{\rho^2}+U_2.
	\end{split}
	\end{equation}
	EVP \eqref{eq:EVP} was solved numerically for a range of $d$ and a couple of values of $\epsilon$, see Fig.~1(a) and discussion in the main text.
	
	For the case $\mathscr{E}_\textsc{dmi}=\mathscr{E}_\textsc{dmi}^{\textsc{b}}$ the energy expression coincides with \eqref{eq:E-2D}, but the DMI term
	\begin{equation}\label{eq:E-2D-B}
	\begin{split}
	\mathcal{E}_{\textsc{2d}}^{\textsc{b}}=&(\vec{\nabla}\theta)^2+\sin^2\theta(\vec{\nabla}\phi)^2+\epsilon\left[\vec{\nabla}\cdot\left(\sin\theta\vec{\eta}\right)\right]^2\\
	&+2d\,\sin^2\theta\left[\vec{\nabla}\theta\times\vec{\eta}\right]_z+4\sin^2\frac{\theta}{2}.
	\end{split}
	\end{equation}
	The corresponding Euler-Lagrange equations coincide with \eqref{eq:theta-phi-2D}, where the replacement $\vec{\eta}\leftrightarrow\bar{\vec{\eta}}$ is made in the DMI term (and only in this term). The second equation is satisfied if $\vec{\eta}=\vec{e}_\chi$, this corresponds to a Bloch skyrmion. The skyrmion profile is determined by the first equation, which in this case coincides with Eq.~(9) with $\epsilon=0$. Thus, $\epsilon$ has no influence on static profiles of the Bloch skyrmion.
	
	The corresponding linearized Landau-Lifshitz equations coincide with \eqref{eq:lin} but the form of the differential operators and potentials
	\begin{equation}\label{eq:H-U-Bloch}
	\begin{split}
	&\hat{\mathrm{H}}_1=-\nabla_\rho^2-\frac{1+\epsilon\cos^2\theta_0}{\rho^2}\partial_{\chi\chi}^2+U_1,\\
	&\hat{\mathrm{H}}_2=-(1+\epsilon)\nabla_\rho^2-\frac{1}{\rho^2}\partial_{\chi\chi}^2+U_2,\\
	&U_1=\frac{\cos2\theta_0}{\rho^2}-d\frac{\sin2\theta_0}{\rho}+\cos\theta_0,\\
	&U_2=\cot\theta_0\nabla_\rho^2\theta_0-\theta_0'^2+\frac{\epsilon}{\rho^2}-d\theta_0',\\
	&W_1=\frac{2+\epsilon}{\rho^2}\cos\theta_0-\frac{d}{\rho}\sin\theta_0,\\ &W_2=W_1+\frac{\epsilon}{\rho}\sin\theta_0\theta_0',\qquad V=\frac{\epsilon}{\rho}\cos\theta_0.
	\end{split}
	\end{equation}
	As for the previous case, the solution of the linear problem is reduced to a generalized EVP, which coincides with \eqref{eq:EVP} but the potentials are determined by \eqref{eq:H-U-Bloch} and the diagonal operators are as follows
	\begin{equation}\label{eq:H-Bloch}
	\begin{split}
	&\hat{\mathscr{H}}_1=-\nabla_\rho^2+\frac{\mu^2(1+\epsilon\cos^2\theta_0)}{\rho^2}+U_1\\
	&\hat{\mathscr{H}}_2=-(1+\epsilon)\nabla_\rho^2+\frac{\mu^2}{\rho^2}+U_2.
	\end{split}
	\end{equation}
	The corresponding EVP \eqref{eq:EVP} was solved numerically for a range of $d$ and a couple of values of $\epsilon$, see Fig.~1(b) and discussion in the main text.

\end{document}